\documentclass[a4,11pt]{article}

\usepackage{graphicx}        
\usepackage[bottom]{footmisc}
\usepackage{comment}
\usepackage{amsmath,amssymb}

\usepackage[authoryear]{natbib}
\usepackage{url}

\usepackage{changepage}

\newenvironment{mydefinition}[1][\unskip]%
{\noindent\rule[2pt]{\textwidth}{0.5pt}\par%
	\noindent\fontfamily{ptm}\selectfont\textbf{#1}}
{\par\noindent\rule[2pt]{\textwidth}{0.5pt}\par}

\newenvironment{digressionAW}%
{\vspace*{0.3\baselineskip}\begin{adjustwidth}{0.03\textwidth}{0.03\textwidth}\fontfamily{ptm}\selectfont\small{\bf Brief digression:}}
{\end{adjustwidth}\vspace*{0.3\baselineskip}}

\begin{document}
	
	\title{Thick Presentism and Newtonian Mechanics}
	\author{Ihor Lubashevsky \\ 
		\small University of Aizu, Ikki-machi, Aizu-Wakamatsu, Fukushima, 965-8580 Japan\\
		\small e-mai: \texttt{i-lubash@u-aizu.ac.jp}}

	\maketitle
	\begin{abstract}
In the present paper I argue that the formalism of Newtonian mechanics stems directly from the general principle to be called the principle of microlevel reducibility which physical systems obey in the realm of classical physics. 
This principle assumes, first, that all the properties of physical systems must be determined by their states at the current moment of time, in a slogan form it is ``only the present matters to physics.'' Second, it postulates that any physical system is nothing but an ensemble of  \textit{structureless} particles arranged in some whose interaction obeys the superposition principle. 
I substantiate this statement and demonstrate directly how the formalism of differential equations, the notion of forces in Newtonian mechanics, the concept of phase space and initial conditions, the principle of least actions, etc. result from the principle of microlevel reducibility. 
The philosophical concept of thick presentism and the introduction of two dimensional time---physical time and meta-time that are mutually independent on infinitesimal scales---are the  the pivot points in these constructions. 
	\end{abstract}
	\tableofcontents

\section{Principle of microlevel reducibility}\label{PMLR}

Dealing with objects of the inanimate world in the frameworks of classical physics, we admit the existence of the microscopic (elementary) level of their description. It means that in modeling such physical systems one can make use of the following premises to be referred further as to the \textit{principle of microlevel reducibility}.%
\footnote{The two premises may be regarded as a particular version of reductionism, a philosophical concept about the relationship between complex systems as whole entities and their constituent parts. In actual fact the concept of reductionism is more complicated and contradictory, for example, there are various versions of reductionism deserving an individual consideration. Nevertheless, it should be noted that the principle of microlevel reducibility may be treated as one of the cornerstones in the research paradigm of physics, namely, it is a particular implementation of the general scientific methods based on decomposition analysis and synthesis \citep[see, e.g.,][]{sep-analysis}.}

\begin{enumerate}
\setlength\itemsep{0.3\baselineskip}
\item \label{PhPost1}
For any physical system there can be found a level of its \textit{microlevel} description at which the system at hand is composed of individual \textit{structureless} entities. The term `structureless' is used to emphasize the fact that either these entities are really structureless or their internal structure does not change in time during analyzed phenomena and so can be treated as a fixed characteristic of the entities.

\item \label{PhPost23} All the properties exhibited by the given system can be explained based on or derived from
\begin{enumerate}
\item \label{PhPost2} the individual properties of these entities which (properties) \textit{exist independently} of the presence of the other entities,

\item \label{PhPost3} the properties of \textit{pairwise} interaction between entities meeting the \textit{superposition principle}.
\end{enumerate}
\end{enumerate}
Further these \textit{structureless} entities will be called particles for short.

Premise~\ref{PhPost3} may be replaced by another one using the concept of fields. Namely, instead of a long-distance interaction of particles a certain field, for example, electromagnetic field is introduced. This field is locally generated by particles, propagates in space, and, in turn, affects them. In these terms Premise~\ref{PhPost3} is read as 
\begin{enumerate}
    \item[2.]  All the properties exhibited by the given system can be explained based on or derived from the individual properties of its constituent structureless particles (item~\ref{PhPost2}) and 
	\begin{enumerate}
		\item[(b$'$)] the own properties of some fields \textit{freely} propagating through space as well as  the properties of the \textit{local} particle-field interaction obeying the \textit{superposition principle} and being responsible for the field generation by the particles and in turn the effects produced by these fields on the particles.
	\end{enumerate}
\end{enumerate}
It should be noted that the concept of particle interaction based on Premise~\ref{PhPost3}$'$ is much reacher in properties and potentiality in describing complex systems in comparison with that based on Premise~\ref{PhPost3}. However, if the dynamics of a certain system is characterized by time scales much longer then the mean time during which the corresponding fields propagate in space over distances about the system size, Premise~\ref{PhPost3}$'$ is approximately reduced to Premise~\ref{PhPost3}. It will be used, for the sake of simplicity, in the following sections, although the results to be obtained can be generalized to theories turning directly to Premise~\ref{PhPost3}$'$. 
Besides, strictly speaking, the use of the fields leads to the necessity of modifying Premise~\ref{PhPost1} too because in this case a given physical system is decomposed not only into structureless particles but also fields existing on their own and which have to be treated as its constituent entities. However various aspects of these fields regarded as individual objects on their own, i.e., beyond the scope of the interaction between the particles that is implemented via these fields, do not belong to the subject-matter of this paper.  

The following two comments are also worthy of noting before passing directly to various consequences of the principle of microlevel reducibility. First, Premise~\ref{PhPost2} concerns the properties that are ascribed to particles individually, i.e. independently of the presence or absence of other particles. In this sentence by the term ``properties'' I actually mean a certain collection of \textit{types} of properties ascribed to the particles individually. For example, ``being located at a spatial point'' is a property type of point-like particles in classical physics, it characterizes a generic feature of all these objects.  ``Being able to restore its previous form once the forces are no longer applied'' exemplifies another generic property which as a type is ascribed to all elastic springs. I noted this fact here to emphasize that particular \textit{instantiations} of these properties, their \textit{tokens}, can depend on the presence of other particles. For example, if a point-like particle $A$ occupies a spatial point $\mathbf{r}$ then another similar particle $B$ cannot be located at this point. Ferroelectricity also exemplifies this feature; in some crystals particular elastic deformation of the crystalline lattice inside a small region can be sustained by similar deformations in other regions via the formation of macroscopic electric field, giving rise to some specific deformation of the crystal as a whole. The difference and relationship between types and their tokens\footnote{A detailed discussion about the distinction between a \textit{type} and its \textit{token} in various aspects can be found, e.g., in the review by \citet{sep-types-tokens}.} is essential for elucidating the basic features of the general scientific methods based on decomposition analysis and synthesis \citep[see, e.g.,][]{sep-analysis}. At the first step,  the individual generic properties of physical particles can be studied dealing with one of them taken separately from the others. After that, at the second step, the complex behavior of ensembles of these particles can be reconstructed based on the found properties and the interaction between the particles which is specified by their spatial arrangement and the particular instantiations of their individual properties.  

The second comment concerns the implementation of this step in reconstructing the behavior of many particle ensembles. In order to do this we need to know how to specify the interaction between the particles. Following the decomposition strategy it could be reasonable to analyze this interaction for a pair of these particles or, at least, a system consisting of a few particles taken separately. In this way, however, we face up to a challenging problem of how the results to be bound can be generalized to the original many particle ensemble. It is solved within the framework of Premise~\ref{PhPost3} appealing to the superposition principle. This principle postulates that the interaction of an arbitrary chosen particle and all the other particles forming a certain ensemble, for example, the cumulative force with which the other particles act on the given one is just the algebraic sum of all the partial forces that can be found in the following way. We should consider a pair of the chosen particle and any one of the other particles assuming the remaining particles of this ensemble to be absent. Then the corresponding partial force is just the force with the second particle of the given pair would act on the first one in this case. In particular, the superposition principle allows us to reduce the interaction energy of a many particle ensemble to the sum of the energies of pair-wise interaction between individual pairs of its particles running over all the possible pairs in this ensemble. Finally, I want to note that Premise~\ref{PhPost3} can be easily generalized to including also plausible three-body forces.


In the next sections I present some arguments about why Newtonian mechanics is based on the mathematical formalism of second order differential equations. At the first step appealing to the principle of microlevel reducibility let us try to elucidate what general mathematical form the laws governing the dynamics of physical systems should have within the framework of classical physics.

\section{Presentism and the time flow}

The possibility of reducing a description of a physical system to \textit{structureless} particles and interaction between them has an important consequence. These particles cannot remember their history or foresee their future; they just have no means to do this, so only the present matters to them. Therefore all the plausible quantities $\{Q\}_\alpha$ that can be used to describe the laws governing the motion of a given particle $\alpha$  have to be taken at the current moment of time. Naturally there should be other characteristics of the particles such as mass, charge, spin magnitude, etc. which, however, are treated as their internal properties not changing in time. Let us regard the dynamics of these particles as their motion in a certain $N$-dimensional space $\mathbb{R}^N$; for our world treated in the realm of classical physics $N=3$. So the spatial position  (spatial coordinates) $x_\alpha$  of the particle $\alpha$  has to enter the collection $\{Q\}_\alpha$. The motion of this particle is represented by the time dependence $x_\alpha(t)$  of its position showing the points occupied previously and the points to be got in future according to prediction of its dynamics. However, for such particles
\begin{itemize}
	\item the past no longer exists,
	\item the future does not exist yet,
	\item only the present matters to them and determines everything.
\end{itemize}   
\noindent
Thereby solely instantaneous characteristics of the particle motion trajectory $\{x_\alpha(t)\}$  may also enter the collection $\{Q\}_\alpha$. They are time derivatives of $x_\alpha(t)$ taken at the current moment of time $t$. In particular, it is the particle velocity $v_\alpha(t) = dx_\alpha(t)/dt$, its acceleration $a_\alpha(t) = d^2x_\alpha(t)/dt^2$, the time derivative of third order called usually the jerk or jolt $j_\alpha(t) = d^3x_\alpha(t)/dt^3$, and so on. However, in order to construct a time derivative we have to consider not only the current position $x_\alpha(t)$ of a particle but also its position $x_\alpha(t-\Delta)$ in the \textit{immediate} past separated from the present by an infinitely short time interval $\Delta\to+0$. Indeed, for example,  the particle velocity is defined as $v_\alpha(t) = \lim_{\Delta\to+0}[x_\alpha(t)-x_\alpha(t-\Delta)]/\Delta$. At this place an attentive reader may find some contradiction, in speaking about the present we actually deal with a certain kind of instants including not only the point-like current moment of time but also other time moments belonging to some neighborhood of the current time whose size may be an infinitely small value. It causes us to speak about the \textit{thick} present. 

The concept of \textit{thick} present is worthy of special attention because it leads directly to the formalism of differential equations and the principle of least action playing a crucial role in modern physics. Therefore let us focus out attention on the philosophical doctrine usually referred to as \textit{presentism} which can be employed to penetrate deeper into the concept of \textit{thick} present.

Broadly speaking, \textit{presentism} is the thesis that only the present exists. In the given form  it is a rather contradictory and ambiguous proposition being one of the subjects of ongoing debates about the nature of time tracing their roots in ancient Greece. In particular, the problems of presentism are met in the famous paradoxes of motion \citep[see, e.g.,][]{sep-paradox-zeno} devised by the Greek philosopher Zeno of Elea (circ. 490--430~BC). Unfortunately, none of Zeno's works has survived and what we know about his paradoxes comes to us indirectly, through paraphrases of them and comments on them, primarily by Aristotle (384--322~BC), however, by Plato (428/427--348/347 BC), Proclus (circ. 410--485 AD), and Simplicius (circ. 490--560 AD) saved them for us. The names of the paradoxes were created by commentators, not by Zeno \citep{dowdenZenoparad}.

We confine ourselves to the \textit{arrow} paradox primarily mentioned in the context of the time problem. This paradox is designed to prove formally that the flying arrow cannot move, it has to be at the rest and, so, the motion is merely an illusion. Citing Aristotle's \textit{Physics}~VI,
\vspace*{0.35\baselineskip}
\begin{adjustwidth*}{0.03\textwidth}{0.03\textwidth}
	[t]he third is \ldots that the flying arrow is at rest, which result follows from the assumption that time is composed of moments \ldots. He says that
	if everything when it occupies an equal space is at rest, and if that which is in locomotion is always occupying such a space at any moment, the flying arrow is therefore motionless.
\end{adjustwidth*}
\vspace*{0.35\baselineskip}
Focusing our attention on the issue in question I want to interpret the arrow paradox as three logical steps:
\begin{itemize}
	\item time is composed of instants---point-like moments of time---and the present is the current moment;
	\item only the present matters, i.e., all the properties of the flying arrow including its motion at a certain velocity are determined completely by its current state, i.e., the spatial point where it is currently located;
	\item whence it follows that the arrow motion is impossible because the state of any arrow---flying from the left to the right, in the opposite direction, or just being at the rest---is the same if at the current moment of time it is located at the same spatial point; the arrow just does not ``know'' in which direction it has to move. 
\end{itemize} 
Aristotle was the first who proposed, in his book \textit{Physics}~VI (Chap.~5, 239b5--32), a certain solution to the arrow paradox. Since that time this paradox having been attacked from various points of view \citep[see, e.g., reviews by][]{lepoidevin2002zeno,sep-paradox-zeno,dowdenZenoparad}, a detailed analysis of Aristotle's solution and its modern interpretation can be found in works by \citet{vlastos1966note,lear1981note,magidor2008another}.  

A na\"ive solution to the arrow paradox could be the proposal to include the instantaneous velocity in the list of basic properties characterizing the current arrow state. Unfortunately the instantaneous velocity, as well as the rate of time changes in any quantity, cannot be attributed to an instant---a point-like moment of time. The velocity is a characteristic of a certain, maybe, infinitesimal neighborhood of this time moment \citep{10.2307/27903678}. For this reason \cite{russell1903principles} rejects the instantaneous velocity at a given moment to be the body's intrinsic property having some causal power. Arguments for and against this view have been analyzed, e.g., by \cite{10.2307/27903678,lange2005can}. 

As a plausible way to overcoming this causation problem of instantaneous velocity, a special version of presentism admits the present to have some duration \citep[e.g.,][]{craig2000extend,dainton2001space,mckinnon2003presentism}. Following \cite{hestevold2008presentism} it is called \textit{thick} presentism. On the contrast, \textit{thin} presentism takes the present to be durationless, which, however, immediately gives rise to logical puzzles like Zeno's arrow. 

In the framework of thick presentism there has been put forward a rather promising solution to the arrow paradox turning to the formalism of nonstandard analysis; for an introduction to this discipline a reader may be referred to \cite{goldblatt2012lectures}. Following  \citep{white1982zeno,mclaughlin1992epistemological,mclaughlin1994resolving,10.2307/27903678,easwaran2014physics,reeder2015zeno} let us equip each point-like time moment $t$ with some neighborhood of infinitesimal thickness $2\epsilon$, i.e., $t\to \mathbf{t}=  (t-\epsilon,t+\epsilon)$ and understand time events as some objects distributed inside $\mathbf{t}$. Here $\epsilon$ is an infinitesimal---infinitely small hyperreal number of nonstandard analysis. Below I will use the term \textit{bold} instants in order to address to such objects and not to mix them with times intervals of finite thickness also conceded in some particular versions of thick presentism. It is worthy of noting that there is no contradiction between the notion of bold instants and the intuitive separability of time moments because for any two moments $t_1$ and $t_2$ separated by arbitrary small but finite interval the infinitesimal regions $\mathbf{t}_1$ and $\mathbf{t}_2$ do not overlap.  

The notion of {bold} instants $\mathbf{t}$ opens, in particular, a gate to endowing the instant velocity with causal power just attributing the instant velocity to the left part $(t-\epsilon,t)$ of $\mathbf{t}$ and assuming that its effect arises in the right part $(t,t+\epsilon)$ \citep[][a similar view was also defended by \citet{lange2002introduction}]{easwaran2014physics}. In this case, as it must, a cause and its effect are ordered in time; a cause precedes its effect.     

Introducing the concept of bold instants we have to accept a special topological connectedness of time which is non-local on infinitesimal scales. Namely, for a time moment $t$ at least all the previous time moments in the infinitesimal interval $(t-\epsilon,t)$ are to coexist, otherwise they cannot have causal power on it. Exactly this connectedness paves the way for properties that can be attributed only to time intervals including infinitesimals to have causal power \citep{lange2002introduction,lange2005can,harrington2011instants}. Allowing the given multitude of time moments to exist we actually accept a special version of thick presentism called the \textit{degree presentism} proposed by \cite{smith2002time}. His account assumes that all events have past and future parts whose existence degree (degree of reality) decreases to zero as their time moments go away from the present. \cite{baron2014priority} has developed a related account of time called \textit{priority presentism} according to which only the present entities exist fundamentally, whereas the past and future entities also existing are grounded in the present. 
 
Any version of presentism has to explain how the flow of time is implemented in dynamical phenomena. In the framework of thick presentism   \cite{baron2012presentism} puts forward the \textit{step-wise} model for the flow of time consisting in temporally extended (thick) instants. Each of these instants comes into and going our of existence in such a manner that successive thick instants partially overlap.

At the next step in describing dynamical processes in terms of thick presentism we face up to a problem of giving the meaning to \textit{time} changes in the properties of some object for which its present partially contains its past and future parts. As a natural way to overcoming this problem, \cite{Smart1949riveroftime} introduces a complex structure of time containing in addition to the \textit{physical} time a certain meta-time. Meta-time is necessary to deal with temporal properties of events embedded into the ``river of time'' when these properties themselves change in time and a meta-time is a place where these changes can occur. 
It should be emphasized that the introduction of two-dimensional time for thick presentism with bold instants does not lead to paradoxes arising in the time travel problem and used often as arguments against the possibility of two-dimensional time structure. A review of these arguments is given, e.g., by \cite{richmond2000plattner,oppy2004can,baron2015back}. The matter is that the difference between the physical time and meta-time becomes essential only within bold instants---the infinitesimal intervals---wherein time travels with non-zero length quantified by standard numbers are impossible. 

Below I will outline my account of dynamical processes consisting in bold instants which is developed for explaining the use of differential equations for modeling dynamical phenomena in classical mechanics and the variational technique as a fundamental law governing dynamics of physical object. Before this, however, let me elucidate the further constructions using the relationship of human and physical time as a characteristic example.  

\begin{figure}[t]
	\begin{center}
		\includegraphics[width=0.9\columnwidth]{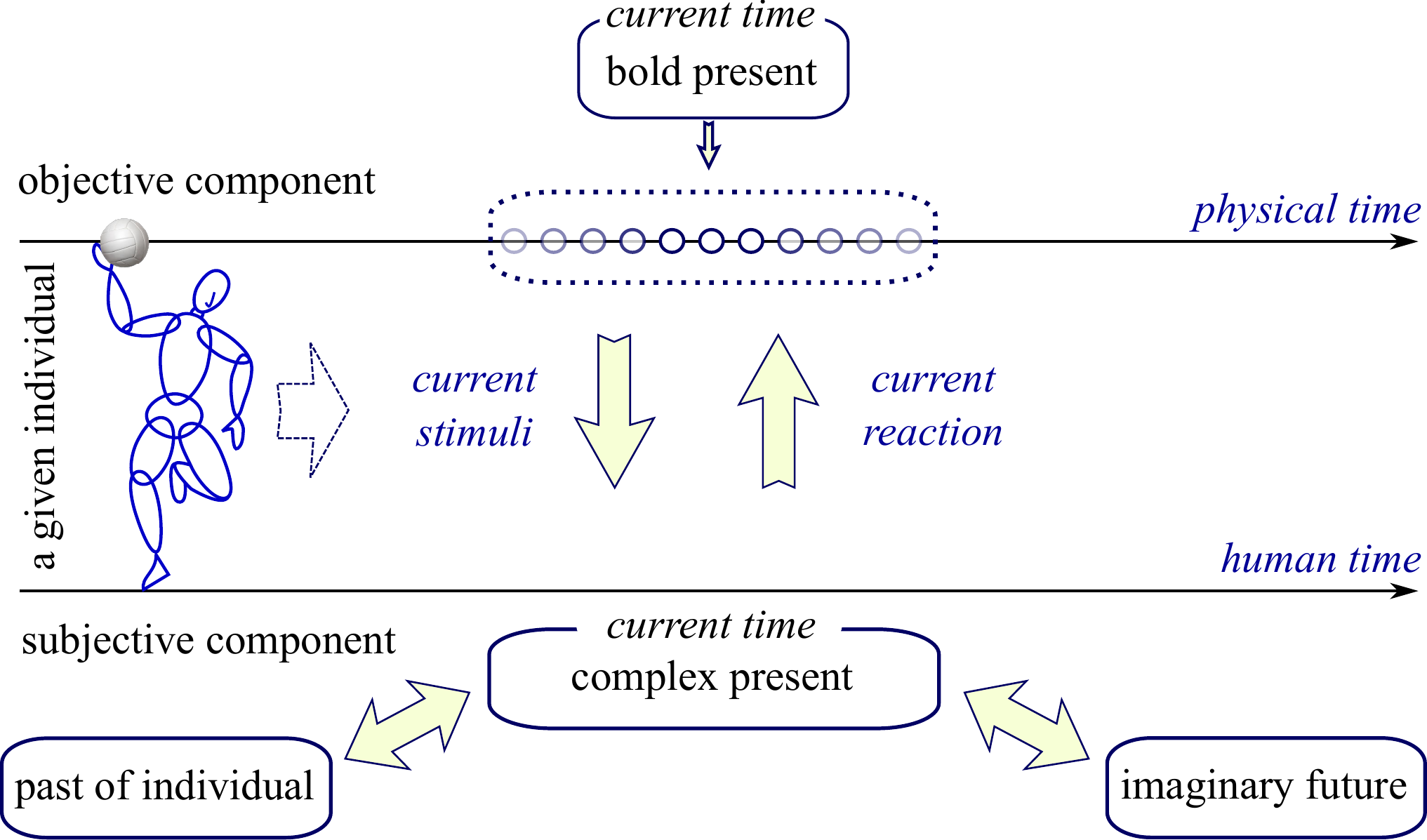}
	\end{center}
	\caption{Illustration of a plausible mechanism synchronizing the time flows in the subjective and objective components of human nature.}
	\label{Fig:Present.1}
\end{figure}

As noted previously \citep{plawinska2009formalism}, two components of human nature, objective and subjective ones (Fig.~\ref{Fig:Present.1}) should be discriminated in modeling human behavior. The objective component represents the world external for a given individual and embedded in the flow of the \textit{physical} time. The subjective component representing the internal world of this individual is equipped with a more complex structure of time to be referred as to \textit{human} time.  It consists of the past retained in the memory, the imaginary future, and the complex (specious) present comprising all the moments of the physical time perceived by the individual as simultaneous. Because the past and future in human mind can affect our current actions we have to regard them as real objects existing in the subjective component. A detail discussion of these temporal components requires immersion in the modern theory of time  which is beyond the capacity of the given paper, so for a review a reader may be referred, e.g., to articles by \citet{sep-time,sep-temporal-parts} and paper-collections edited by \citet{callender2011oxford,ciuni2013present}. Here we touch only the problem how these time components are related to each other, moreover, confine ourselves to the nearest past, the complex present, and the forthcoming future.  

Generally speaking we may say that the complex present in the human time is synchronized with the present in the physical time  via the direct interaction of the given individual with the reality. However this synchronization is not an one-to-one map. In fact, the complex present may be conceived of as a certain interval $\mathfrak{T}_t$ with fuzzy boundaries containing the current moment $t$ of the physical time. Its any point $t'$ is perceived by the individual as the present in the human time with some degree $\Phi(t-t')$ decreasing to zero as the time gap $|t-t'|$ increases and exceeds the characteristic duration $\Delta$ of the complex present. It is necessary to emphasize that on scales about $\Delta$ the order of time moments in the physical time is not recognized by the given individual and so does not exist in the subjective component. 

As far as the nearest past in the subjective component is concerned, it may be regarded as fixed. On the contrast, the forthcoming imaginary future permanently changes as its time moments $t'$ come closer to the present, $t'\to \mathfrak{T}_t$, and becomes the fixed reality when the point $t'$ goes into $\mathfrak{T}_t$. It is a result of permanent correction of the imaginary future based on the interaction between the individual and the reality.      

The given example prompts me to put forward the following model of the time flow consisting in bold instants applied to describing dynamics of a certain physical system. 

\section{Thick presentism with moving window of existence}

The non-stand analyses enables us to operate with infinitely small and infinitely large numbers in addition to standard ones. The set of these numbers forms a field, i.e., all the arithmetics operations (addition, subtraction, multiplication, and devision), relations, and, thus, many mathematical functions are defined in it \citep[for an introductin see, e.g.,][]{goldblatt2012lectures}. It allows us to deal with infinitesimal bold instants as ordinary intervals. 

\begin{figure}[t]
	\begin{center}
		\includegraphics[width=1\columnwidth]{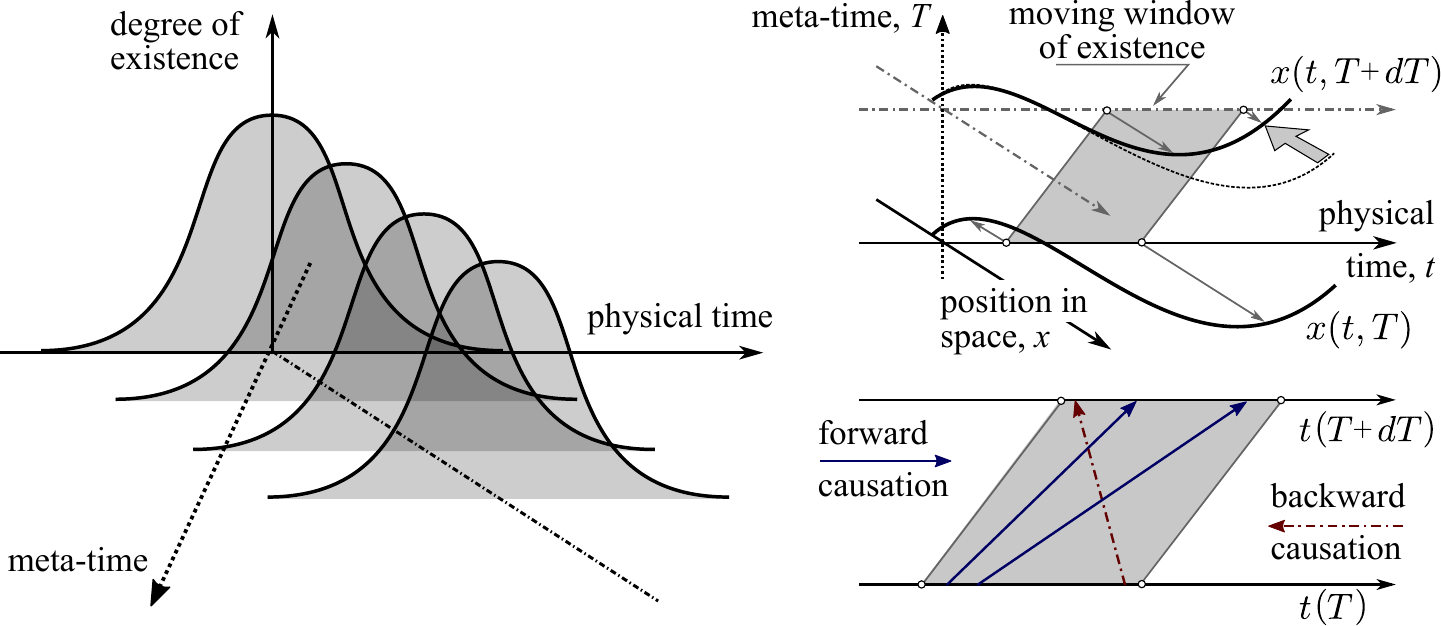}
	\end{center}
	\caption{Illustration of thick presentism with moving window of existence: (left) two-dimensional time structure and (right) the realization dynamics of some physical system.}  
	\label{Fig:Present.2}
\end{figure}

Using infinitesimals we can introduce a certain function $\Phi(|t'-T|/\epsilon)$ giving us the degree of existence for the events coming into being at a time moment $t'$. Here $\epsilon$ is the infinitesimal thickness of the bold instant $\mathbf{t}$ centered at $T$ and $\Phi(|t'-T|/\epsilon)\to 0$ as the ratio $|t'-T|/\epsilon\to\infty$.
The function $\Phi(|t'-T|/\epsilon)$ admits the interpretation as a certain characteristic function of the \textit{window of existence} with fuzzy boundaries which is attributed to the bold instant $\mathbf{t}$. This window moves along the axis of the \textit{physical} time due to the \textit{flow of time}. Actually these constructions introduce a two dimensional (2D) time structure containing the meta-time governing the realization of physical systems on the corresponding time-space manifold (Fig.~\ref{Fig:Present.2}). 

Having introduced the bold instants---time \textit{intervals} even if they are of infinitesimal thickness---as the basic elements of time flow we have to modify the standard way of describing the dynamics of a certain system in the space $\mathbb{R}^N$. Within the standard description the system is specified by the point-like position $x(t)$ it occupies at the current moment of time $t$. As time grows the generated trajectory represents the system motion. In the realm of thick presentism we should ascribe a certain degree of existence not only to the point-like object $x(t)$ but also to the trajectory fragments $\{x(t)\}_{\mathbf{t}}$, where $t\in\mathbf{t}$. It means that the very basic level of the system description must consist in the trajectories, at least, their parts rather than point-like objects and causal power may be attributed only to these basic elements. We should to do this at each  moment $T$ of meta-time, otherwise, evolution and emergence as dynamical phenomena are merely a mirage---everything is fixed beforehand. In other words, the basic element of the system description in the 2D-time structure is given by the trajectory $\{x(t,T)\}$ or, speaking more strictly, its partition specified by bold instants $\mathbf{t}$. I have used the term \textit{trajectory} to emphasize that these basic elements are certain functions of the argument $t$ running from $-\infty$ to $+\infty$ rather than points of the space $\mathbb{R}^N$; here the meta-time $T$ plays the role of a parameter. 

In these terms the system dynamics at any moment $T$ of meta-time is characterized by the following components: 
\begin{itemize}
    \item the past of the system: $\{x(t,T)\}$ matching $t<T$ and $t\notin\mathbf{t}_T$,
    \item the thick present of the system: $\{x(t,T)\}$ where $t\in\mathbf{t}_T$,
    \item the future of the system: $\{x(t,T)\}$ matching $t>T$ and $t\notin\mathbf{t}_T$
\end{itemize}    
depending on $T$. It is worthy of noting that involving the past and further into consideration of causal processes affecting the system  dynamics does not contradict the previous statement about their absence for structureless particles of classical physics. Such particles have no means to remember \textit{individually} their past or to predict their future. However in the case under consideration the causal power of the past and future is due to the physical properties of the  time flow itself rather than that of the particles and spends over temporal intervals of infinitesimal thickness only.   

In the framework of thick presentism all the properties of the given system at the current moment $T$ of meta-time must be determined completely by the trajectory $\{x(t,T)\}$, whereas the presence of its points in the reality is determined by the current position of the window of existence. It concerns also the property I call the \textit{sensitivity} of the given system to the flow of meta-time or simply \textit{meta-time sensitivity}. It quantifies the variation of the trajectory $\{x(t,T)\}$ \textit{caused} by the meta-time flow provided the corresponding part of the trajectory is present in the reality.  The partial existence of a trajectory fragment in the reality decreases its variation so the governing equation for these trajectory variations can be written as 
\begin{equation}\label{present:1}
\frac{\partial x(t,T)}{\partial T} = P\left(\frac{|t-T|}{\epsilon}\right)\widehat{\Omega}\left[\{x(t,T)\}\right],
\end{equation}
where the operator $\widehat{\Omega}\left[\{x(t,T)\}\right]$ specifies the meta-time sensitivity of the given system with the trajectory $\{x(t,T)\}$. Figure~\ref{Fig:Present.2} (right fragment) illustrates the variations of the system trajectory as the meta-time grows. It should be noted that within the bold instant $\mathbf{t}_T$ the time moments may not be ordered in their effects, i.e., the variation of the system trajectory at moment $t\in\mathbf{t}_T$ can be partially caused by time moments preceding as well as succeeding it. In the latter case we can speak about backward causation \cite[for a general discussion on the backward causation problem a reader may be referred to][]{sep-causation-backwards}.

\begin{figure}[t]
	\begin{center}
		\includegraphics[width=0.8\columnwidth]{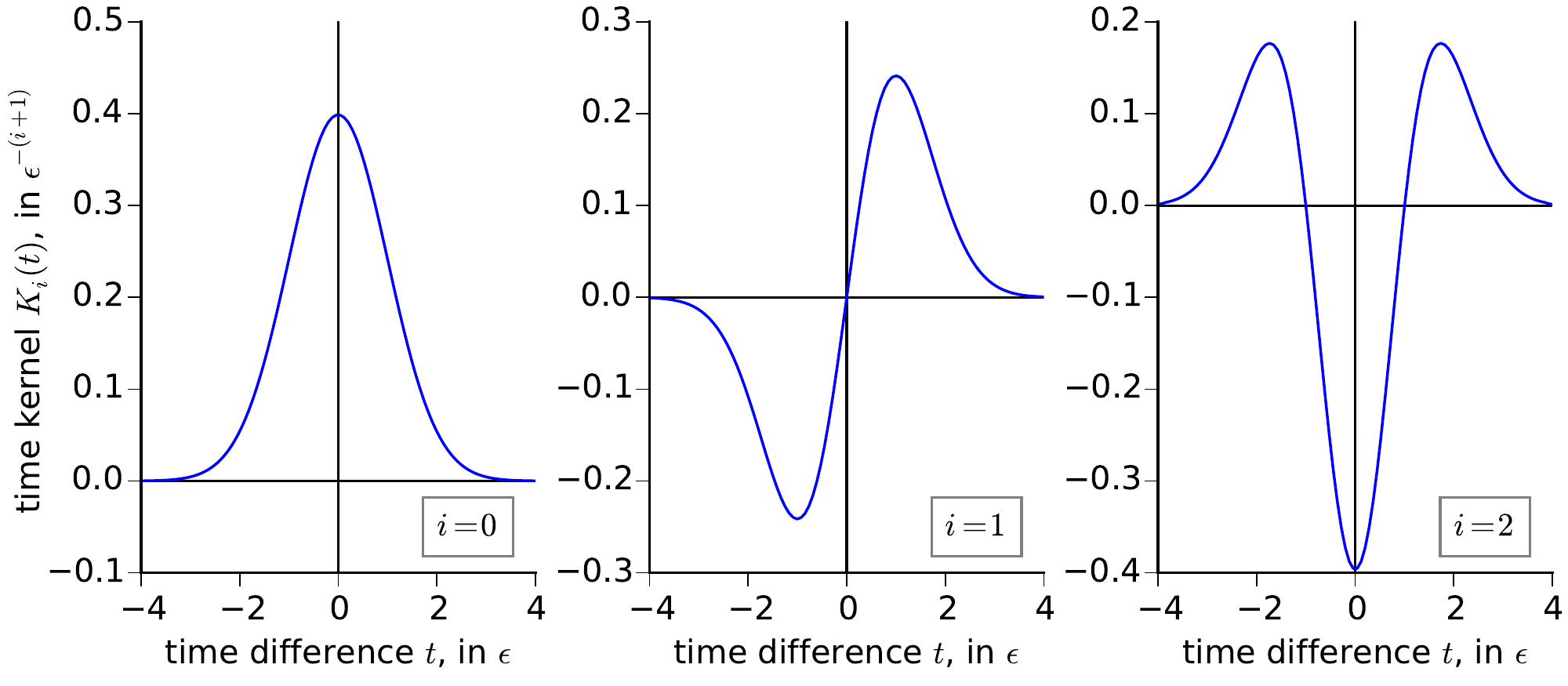}
	\end{center}
	\caption{Typical forms of kernels determining nonlocal contribution of different time moments to the system meta-time sensitivity.}
	\label{Fig:Present.3}
\end{figure}

Equation~\eqref{present:1} can relate to one another only the trajectory fragments corresponding to bold instants $\mathbf{t}$ that either contain  the time moment $T$ or are distant from it over scales about $\epsilon$. So terms similar to
\begin{equation}\label{present:coff}
	 \int\limits_{-\infty}^{+\infty}dt'\, K_i\left(\frac{t-t'}{\epsilon}\right)x(t',T)
\end{equation}
should mainly contribute to the variation of the trajectory $\{x(t,T)\}$ at the point $t\in\mathbf{t}_T$ and the typical forms of the kernels $ K_i\left(\ldots\right)$  are exemplified in Fig.~\ref{Fig:Present.3}. Such nonlocal effects can connect only time moments separated by infinitely small time lags whereas the motion trajectory of systems at hand are to be smooth curves. In this case the nonlocal operator $\widehat{\Omega}\left[\{x(t,T)\}\right]$ should reduce to a certain local function $\omega$ whose arguments are the current system position $x$ and its various derivatives taken at the current moment $t$
\begin{equation}\label{present:2}
	\widehat{\Omega}\left[\{x(t,T)\}\right] \Longrightarrow \omega\left[x(t,T),\frac{\partial x(t,T)}{\partial t}
	,\frac{\partial^2 x(t,T)}{\partial t^2}
	,\frac{\partial^3 x(t,T)}{\partial t^3},\ldots
	\right].
\end{equation}
The possible forms of this function and the corresponding consequences will be discussed in the following two sections. In the remaining part of this section I will explain the mechanism via which \textit{steady-state} laws governing system dynamics can emerge in the realm of thick presentism.

The window of existence moves from past to future along both the time-axes and never returns to instants already passed. So in the realm of thick presentism the past of a given system cannot change as the meta-time grows, which stems directly from equation~\eqref{present:1}. However it does not mean that the future has no influence on the past. In the general case the past is formed  during the trajectory transformation at time moments when the window of reality passes through them and the result depends on the ``initial'' details of the system trajectory $\{x(t,T)\}$ in the region $t>T$. In this case it is not possible to speak about universal laws governing the system dynamics in the standard interpretation. Nevertheless there a special case when it becomes possible.

\begin{figure}[t]
	\begin{minipage}[b]{0.6\columnwidth}
		\includegraphics[width=\columnwidth]{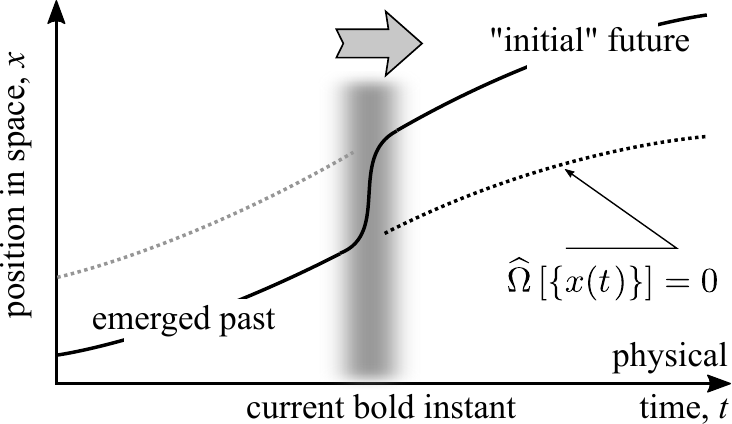}
	\end{minipage}\qquad
	\begin{minipage}[b]{0.34\columnwidth}
		\caption{Emergence of the system past matching the system dynamics governed by stable laws which are determined completely by the system physics.}
		\label{Fig:Present.4}
	\end{minipage}
\end{figure} 

According to equation~\eqref{present:1} the change of the system trajectory $\{x(t,T)\}$ is characterized by two temporal scales. The first one is the thickness of bold instants, $\epsilon$, specifying the duration of the time interval within which a given point of the system trajectory is in present. The second one is the time scale $\tau$ characterizing the rate of the conversion of forthcoming future into the nearest past within the current bold instant $\mathbf{t}_T$, i.e., the strength of the operator $\widehat{\Omega}\left[\{x(t,T)\}\right]$.  When the ration 
\begin{equation}\label{present:lim}
\frac{\tau}{\epsilon}\ll 1\quad\text{or, moreover, is itself infinitesimal,}
\end{equation}
the system trajectory gets equilibrium configuration (if it is stable) actually within the bold instant $\mathbf{t}_T$ which is described by the condition
\begin{equation}\label{present:3}
	\widehat{\Omega}\left[\{x(t)\}\right] = 0\,,
\end{equation}    
where the \textit{steady-state} trajectory $\{x(t)\}$ does not depend on meta-time $T$. 

In this case the system past is mainly determined by equality~\eqref{present:3} and ``forgets'' completely the ``initial'' future of the system. As a result the laws describing the newly emerged past as the present may be of a universal form reflecting only the physics of a given system. Figure~\ref{Fig:Present.4} illustrates this situation.

\section{Steady-state laws of system dynamics}\label{SSLSD}

I will call condition~\eqref{present:lim} the limit of \textit{steady-state laws} and will assume it to hold in our reality. In this case the system dynamics described in terms of the position $x(t)$ in the space $\mathbb{R}^N$ occupied by the system in the \textit{immediate} past obeys the equation
\begin{equation}\label{present:10}
\omega\left[x(t),\frac{d x(t)}{d t}
,\frac{d^2 x(t)}{d t^2}
,\frac{d^3 x(t)}{d t^3},\ldots
\right] = 0
\end{equation}
by virtue of \eqref{present:2}.  

Expression~\eqref{present:10} is the key point determining further constructions in the present paper. In particular, it explains why the laws governing the dynamics of systems in the framework of classical physics admits a representation in the form of some formulas joining together the time derivatives of the motion trajectory taken at the current moment of time. Thereby the formalism of differential equations is actually the native language of physics or, speaking more strictly, Newtonian mechanics. Naturally the question on whether differential equations are the very basic formalism of physics has been in the focus of long-term debates and attacked from various points of view, for a short review see, e.g., \citet{stoltzner2006classical} and references therein.

Causal relations can be also attributed to law~\eqref{present:10}, at least, when the list of arguments of the function $\omega(\ldots)$ is finite. In this case resolving equation~\eqref{present:10} with respect to the highest order $m$ time derivative of $x(t)$ we obtain the expression
\begin{equation}\label{present:11}
\frac{d^m x}{dt^m} =\Phi \left(
x,\,\frac{dx}{dt},\,
\frac{d^2x}{dt^2},\,\ldots\,,\frac{d^{m-1}x}{dt^{m-1}}
\right)
\end{equation}
which admits interpretation as a causal relationship between the lower order time derivatives 
\begin{equation}\label{present:12}
  x,\,\frac{dx}{dt},\,
  \frac{d^2x}{dt^2},\,\ldots\,,\frac{d^{m-1}x}{dt^{m-1}}
\end{equation}
playing the role of causes and the highest time derivative ${d^m x}/{dt^m}$ being their effect. Indeed, when the system trajectory undergoes sharp variations inside the bold instant $\mathbf{t}_T$ the highest derivative demonstrates changes most drastically and it is possible to say that the collection of quantities~\eqref{present:12} finally cause the highest derivative to take value~\eqref{present:11}. 

The accepted hypothesis on the finite number of arguments in the function $\omega(\ldots)$ can be directly justified if the field of hyperreal numbers is extended to a ring including nilpotent infinitesimals. Nilpotents are nonzero infinitely small numbers that yield zero when being multiplied by themselves for a certain number of times. So if the kernels $K_i(\ldots)$ contain nilpotent cofactors, the meta-time sensitivity operator $\widehat{\Omega}[\{x(t,T)\}]$ can comprise only finite order power terms with respect to quantities similar to \eqref{present:coff}. In a similar way \citet{reeder2015zeno} uses nilpotents for constructing a novel solution to Zeno's arrow.  

\section{Variational formulation of steady-state dynamics}\label{VFSSD}

There is a special case worthy of individual attention that admits the introduction of a certain functional
\[
   \mathcal{L}[\{x(t,T)\}]
\]
to be call \textit{action} following the traditions accepted in physics. This functional specifies the operator of meta-time sensitivity as its functional derivative
\begin{equation}\label{present:var1}
 \widehat{\Omega}[\{x(t,T)\}] = -\frac{\delta S[\{x(t,T)\}]}{\delta x(t,T)}\,.
\end{equation}
Because bold instants can couple only infinitely close time moments the action functional in the general form can be written as
\begin{equation}\label{present:var2}
    \mathcal{L}[\{x(t,T)\}] = \int\limits_{-\infty}^{+\infty}dt\, 
    L\left[x(t,T),\frac{\partial x(t,T)}{\partial t}
    ,\frac{\partial^2 x(t,T)}{\partial t^2}
    ,\frac{\partial^3 x(t,T)}{\partial t^3},\ldots
    \right],
\end{equation} 
where function
\begin{equation}\label{present:var3}
L\left[x(t,T),\frac{\partial x(t,T)}{\partial t}
,\frac{\partial^2 x(t,T)}{\partial t^2}
,\frac{\partial^3 x(t,T)}{\partial t^3},\ldots
\right]
\end{equation}
is called the Lagrangian of a given system.

In the limit of stead-state laws the trajectory $\{x(t)\}$ meeting condition~\eqref{present:3} should be stable with respect to small (infinitesimal) variations  
\[
x(t,T) = x(t) + \delta x(t,T)\,.
\]
It means that the variations $\delta x(t,T)$ have to fade as meta-time $T$ grows. This stability condition directly gives rise to the following requirement which has to be imposed on the corresponding form of the action functional and its Lagrangian. 

\begin{mydefinition}[Principle of Least Actions:]
Let a physical system admit the introduction of the action functional \eqref{present:var2} describing its dynamics in meta-time. Then its steady-state trajectory $\{x(t)\}$ describing the system motion in the past including the immediate past matches the minimal value of the action functional among all the other possible trajectories
\begin{equation}\label{present:var4}
x(t) \Longrightarrow \min  \int\limits_{-\infty}^{+\infty}dt \,
L\left[x(t),\frac{dx(t)}{d t}
,\frac{d^2 x(t)}{d t^2}
,\frac{d^3 x(t)}{d t^3},\ldots
\right].
\end{equation}

\end{mydefinition}
\noindent
Actually this principle is in one-to-one correspondence with the principle of least actions well-known in physics provided the Lagrangian $L(x,dx/dt)$ depends only on the system position $x$ and the velocity $dx/dt$.

\section{Notion of phase space}\label{FDENPS}

In the previous sections we have considered the general description of system dynamics in the framework of thick presentism and the limit of steady-state laws has been assumed to hold in our world. In this case the dynamics of a physical system conceived of as the motion of a point $x$ in a certain space is governed by equation~\eqref{present:10} joining together all the time derivatives of the trajectory $x(t)$ taken at the current moment of time $t$.

In what follows, first, we will confine ourselves to the case where the number of the time derivatives entering the right-hand side of \eqref{present:10} is finite for any physical object. Second, we will consider an ensemble of structureless particles whose individual motion can be represented as the motion of a point $x_\alpha$ in the space $\mathbb{R}^N$; in our world $N=3$. This ensemble may be described as a point $x=\{x_\alpha\}$ of the space $\mathbb{R}^{NM}$, where $M$ is the number of particles in the given ensemble.  Besides, for the sake of simplicity we will assume that for all the particles only the first $(m-1)$ derivatives of their coordinates $x_\alpha$ enter equation~\eqref{present:10}.\footnote{The further constructions can be easily generalized to the case when the state of different particles is characterized by different parameters $m_\alpha$, which, however, over-complicates the mathematical expressions without any reason required for understanding the subject.} 

Under these conditions equation~\eqref{present:10} treated as some equality can be reserved with respect to the highest derivative $d^m x/dt^m$ which gives us expression~\eqref{present:11}. This expression may be interpreted as a causal type relationship between the time derivatives of order less than $m$ (including the zero-th order derivative just being the particle positions) and the derivative $d^m x/dt^m$. For individual particles formula~\eqref{present:11} takes the form
\begin{equation}\label{Red:2}
\frac{d^m x_\alpha}{dt^m} =\Phi_{\alpha} \left(
x,\,\frac{dx}{dt},\,
\frac{d^2x}{dt^2},\,\ldots\,,\frac{d^{m-1}x}{dt^{m-1}}
\right)\,,
\end{equation}
where the particle index $\alpha$ is omitted at the list of arguments in the right-hand side of \eqref{Red:2}, which denotes that all the particles of a given ensemble should be counted here because of the particle interaction.  

The fact that the right-hand side of equation~\eqref{Red:2} contains only the time derivatives of order less than $m$ does not mean the mutual independence of these quantities. There could be conceived of some additional constrains imposed on this system such that one of these derivatives, mainly, $d^{m-1} x/dt^{m-1}$ is completely determined by the others. It actually reduces the number of arguments in \eqref{Red:2}. Therefore below we may assume the collection of quantities
\begin{equation}\label{Red:1}
\{Q\}_\alpha = \left\{
x,\,\frac{dx}{dt},\,
\frac{d^2x}{dt^2},\,\ldots\,,\frac{d^{m-1}x}{dt^{m-1}}
\right\}_\alpha,
\end{equation}
to be \textit{mutually independent} for all the particles $\{\alpha\}$. It means that for arbitrary chosen values there can be found an instantiation of this system such that during its motion these time derivatives take the given values at a given moment of time $t$.   

\begin{figure}
	\begin{center}
		\includegraphics[width=0.95\columnwidth]{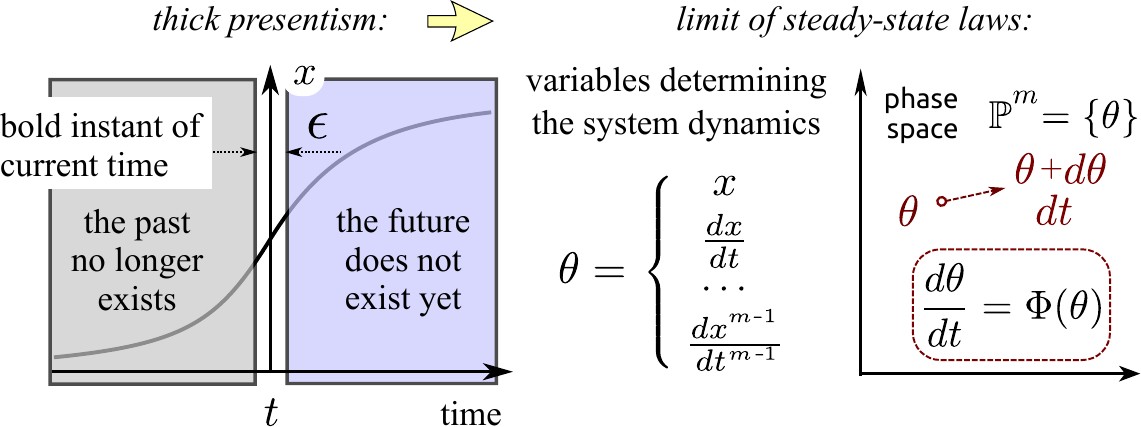}
	\end{center}
	\caption{Illustration of the phase space introduction starting from the principle of microlevel reducibility in the framework of thick presentism.}
	\label{Fig:Reduct}
\end{figure}

Now we can introduce the notion of the \textit{phase space} 
\begin{equation}\label{Red:3}
\mathbb{P} = 
\left\{
x,\,\frac{dx}{dt},\,
\frac{d^2x}{dt^2},\,\ldots\,,\frac{d^{m-1}x}{dt^{m-1}}
\right\}
\end{equation}
for the system at hand regarded as a whole. If we know the position of the system in the space $\mathbb{P}$ treated as a point $\theta$ with the coordinates
\begin{equation}\label{Red:4}
\theta =
\left(
x,\, \frac{dx}{dt},\,
\frac{d^2x}{dt^2},\,\ldots\,,\frac{d^{m-1}x}{dt^{m-1}}
\right),
\end{equation}
then the rate of the system motion in the phase space $\mathbb{P}$  is completely determined via relationship~\eqref{Red:2}. Solving this equation we can construct the trajectory of the system motion. The aforesaid is illustrated in Fig.~\ref{Fig:Reduct}.

The phase space is one of the basic elements in describing such objects. In particular, specifying the system position in the phase space $\mathbb{P}= \{\theta\}$  we actually can calculate the velocity of the system motion in it. Indeed, the time derivatives entering the complete collection of mutually independent components for all the particles
\[
\Bigg\{
\left\{
x,\,\frac{dx}{dt},\,
\frac{d^2x}{dt^2},\,\ldots\,,\frac{d^{m-1}x}{dt^{m-1}}
\right\}_\alpha\Bigg\}
\]
may be treated as independent phase variables
\[
\varphi_{\alpha,p} \,\overset{\text{def}}{=}\, \frac{dx_\alpha^{p-1}}{dt^{p-1}}\,,\qquad\text{for $p=1,2,\ldots, m$},
\]
in particular, the variable $\varphi_{\alpha,1}=x_\alpha$  just represents the spacial coordinates of the particle $\alpha$, $\varphi_{\alpha,2}=v_\alpha$ is the velocity of its motion in the physical space, and $\varphi_{\alpha,3}=a_\alpha$ is its acceleration in it. The mutual independence of these variables is understood in the sense explained above; it is the principle possibility of finding a real instantiation of the system in issue at the state such that at a given moment of time all these derivatives take the corresponding values chosen arbitrary. Then the system dynamics governed by equation~\eqref{Red:2} may be represented as
\begin{subequations}\label{Red:5}
	\begin{align}
	\nonumber
	\frac{d\varphi_{\alpha, p}}{dt} & = \varphi_{\alpha,p+1} \quad\quad\text{for $p=1,2, \ldots, m-1$ and}\\
	\label{Red:5a}
	\frac{d\varphi_{\alpha, m}}{dt} & = \Phi_{\alpha} \left(\varphi_1, \varphi_2, \ldots,\varphi_m \right).
	\end{align}
	Here, as previously, omitting the index $\alpha$ at the arguments of the function $\Phi_\alpha\{\ldots\}$ denotes that its list of arguments should contain the phase variables $\{\varphi_{\alpha,p}\}$ of all the particles. These differential equations which symbolically may be written as
	\begin{equation}\label{Red:5b}
	\frac{d\theta}{dt} = \Phi(\theta)
	\end{equation}
\end{subequations}
determine all the laws of the system dynamics.

The existence of equations~\eqref{Red:5} endows the inanimate world in the realm of classic physic with a fundamental property described by two notions reflecting its different aspects.  One of them is the notion of \textit{initial conditions}.\footnote{Actually the range of applicability of notion of initial conditions is much wider than Newtonian mechanics, which however is beyond the scope of our discussion.} Namely, if we know the system position $\theta_0$  in the phase space $\mathbb{P}$  at an \textit{arbitrary chosen} moment of time $t_0$  then, generally speaking, equations~\eqref{Red:5} possess the unique solution for $t>t_0$ 
\begin{equation}\label{Red:6}
\theta  = \theta(t,\theta_0)\quad
\text{such that at $t=t_0$}\quad \theta(t_0,\theta_0) = \theta_0\,.
\end{equation}
In other words, if the ``forces'' $\Phi(\theta)$  and the initial system position $\theta_0$ are known, then the system dynamics can be calculated, at least, in principle. It means that inanimate systems have no memory; if we know what is going on with such a system at a given moment of time, then its ``history'' does not matter, which was claimed previously appealing to Premise~\ref{PhPost1}.

The other one is the notion of the \textit{determinism} of physical systems; if we repeat the system motion under the same conditions with respect to the initial position $\theta_0$  and the ``forces'' $\Phi(\theta)$ acting on the system, then the same trajectory of system motion will be reproduced. Drawing this conclusion we actually have assumed implicitly that the ``forces'' $\Phi(\theta)$ do not depend on the time $t$. If it is not so, then we can expand the system to include external objects causing the time dependence of these ``forces.'' The feasibility of such an extension is justified by the principle of microlevel reducibility. In fact it claims that at the  microslevel describing completely a given system there are only structureless constituent particles and the interaction between them. So there no factors that can cause the time dependence of the ``forces'' $\Phi(\theta)$ and, in particular, endow them with random properties.

\begin{digressionAW}
	It should be noted that this determinism does not exclude highly complex dynamics of nonlinear physical systems manifesting itself in phenomena usually referred to as dynamical chaos. Dynamical chaos can be observed when the  motion of a system in its phase space is confined to a certain bounded domain and the motion trajectories are unstable with respect to small perturbations. This instability means that two trajectories of such a system initially going in close proximity to each other diverge substantially as time goes on, and, finally, the initial proximity of the two trajectories becomes unrecognizable. These effects make the dynamics of such systems \textit{practically} unpredictable. For example, in numerical solution of equations~\eqref{Red:5} the discretization of continuous functions and round-off errors play the role of disturbing factors responsible for a significant dependence of the found solutions on the selected time step in discretization and particular details of arithmetic operations at a used computer. In studying systems with dynamical chaos in laboratory experiments the presence of weak uncontrollable factors is also inevitable. Moreover, there is a reason arguing for the fact that the notion of dynamical chaos is a fundamental problem rather than a particular question about practical implementations of system dynamics. The determinism of physical systems implies the reproducibility of their motion trajectories provided the \textit{same} initial conditions are reproduced each time. However, in trying to control extremely small variations in the system phase variables we can face up to effects lying beyond the range of applicability of classical physic. So in studding various instantiations of one system it can be necessary to assume that each time the initial conditions are not set equal but distributed randomly inside a certain, maybe, very small domain. So determinism and dynamical chaos are not contradictory but complementary concepts reflecting different aspects of the dynamics of physical systems in the realm of classical physics. 
\end{digressionAW}

\section{Energy conservation and Newton's second law}\label{NewMech}

Appealing to the concepts of thick presentism it is not possible to find out the order $m-1$  of the derivative $d^{m-1}x/dt^{m-1}$ that determines how many components collection~\eqref{Red:1} contains, i.e. to specify the structure of the phase space $\mathbb{P}$~\eqref{Red:3}. From physics we know that $m=2$, i.e.,  for any ensemble of classical particles  the phase space  consists of the spatial coordinates and velocities of the particles making up it. Let us try to elucidate whether this type phase space endows the corresponding systems with unique properties via which such systems stand out against the other objects.

In the simplest case, i.e., when the value $m=1$, the phase space contains only the spatial positions of particles $\mathbb{P}^1 = \{x\}$. In this instance the respective systems tend to go directly to spacial ``stationary'' points $x_\text{eq}$  such that 
\begin{equation*} 
\frac{d x_\alpha}{dt} =\Phi_{\alpha,1} \left(x_\text{eq} \right) = 0\quad\text{for all $\alpha$,}
\end{equation*}
if, naturally, they are stable. This class of models, broadly speaking, is the heart of Aristotelian physics assuming, in particular, that for a body to move some force should act on it. There are many examples of real physical objects exhibiting complex behaviour that are \textit{effectively} described using the notions inherited from Aristotelian physic. The complexity of their dynamics is due to the fact that all their stationary points turn out to be unstable and, instead, some complex attractors, i.e., multitudes toward which systems tend to evolve, arise in the phase space $\mathbb{P}^1$. Nevertheless, if our inanimate world were governed solely by  Aristotelian physics it would be rather poor in properties. For example, if the motion of planets of a solar system obeyed such laws then they would drop to its sun and the galaxies could not form. 

The next case with respect to the simplicity of phase spaces matches $m=2$. It is our world; the phase space of physical particles, at least, within Newtonian mechanics consists of their spatial coordinates and velocities, 
\begin{equation}\label{Red:7}
\mathbb{P}^2=\left\{x,v=\frac{dx}{dt}\right\},
\end{equation}
which together determine the next order time derivative, the particle acceleration, 
\begin{equation} \label{Red:6empty:1}
a_\alpha=\frac{d^2 x_\alpha}{dt^2} =\Phi_{\alpha,2} \left(x,v \right).
\end{equation}
In other words, in Newtonian physics for a body to \textit{accelerate} some forces should act on it, whereas in  Aristotelian physics for a body to \textit{move} some forces should act on it. The systems whose dynamics is described by the phase space $\mathbb{P}^2$ possess two distinctive features.

One of them, usually called the dynamics reversibility, is exhibited by systems where the regular ``force'' $\Phi_{\alpha,2} \left(x\right)$  depends only on the particle positions $\{x\}$. In this case the governing equation~\eqref{Red:6empty:1} is symmetrical with respect to changing the time flow direction, i.e. the replacement $t\to -t$. This symmetry is responsible for the fact that if at the end of motion the velocities of all the particles are inverted, $v_\alpha\to-v_\alpha$, then they should move back along the same trajectories. 

The other one is the possibility of introducing the notion of energy for the real physical systems. At the microlevel the energy, comprising the components of the kinetic and potential energy, is a certain function
\begin{equation}\label{Red:9}
\mathcal{H}(x,v)
\end{equation}
whose value does not change during the system motion. Namely, if $x(t)$ is a trajectory of system motion then the formal function on $t$
\begin{equation}\label{Red:10}
H(t)\,\overset{\text{def}}{{}={}}\, \mathcal{H}\left[x(t), v(t)=\frac{dx(t)}{dt}\right]
\end{equation}
in fact does not depend on the time $t$. Such systems are called conservative. The existence of the energy $\mathcal{H}(x,v)$ does not necessary stem from the governing equation~\eqref{Red:6empty:1} but is actually an additional assumption about the basic properties of physical systems at the microlevel. Naturally, it imposes some conditions on the possible forms of the function $\Phi_{\alpha, 2}(x,v)$. 

The two features endow physical systems with rich properties and complex behavior. For example,  although in a solar system the planets are attracted by the sun, they do not drop on it because when a planet comes closer to the sun its kinetic energy grows, preventing the direct fall on the sun. Naturally, this planet should not move initially along a straight line passing exactly through the sun. The reversibility is responsible for this planet to tend to return to the initial state or its analogy after passing the point at the planet trajectory located at the shortest distance to the sun. Broadly speaking, the existence of energy endows physical objects with a certain analogy of memory. Certainly, if the initial conditions for a given system are known, its further dynamics is determined completely, at least, in principle, so the previous system history does not matter. Nevertheless, the conservative systems ``do not forget'' their initial states in the meaning that the motion trajectories matching different values of the energy cannot be mixed.%
\footnote{First, it should be noted that a many-particle ensemble can exhibit so complex dynamics that it could be impossible to track its motion from a given initial state within physically achievable accuracy. In this case it possible to speak about the effective forgetting of the initial conditions. The latter also concerns extremely weak perturbations.
Second, there are systems with highly complex dynamics whose description does not admit any energy conservation and their motion is irreversible; the term dynamical chaos noted before is usually used to refer to these phenomena. Nevertheless it does not contradict to the present argumentation because the corresponding irreversible description is obtained via the reduction of equation~\eqref{Red:6empty:1} and assuming the presence of a certain external environment weakly interacting with a system at hand.}

Summarizing this discussion about the systems with the phase space $\mathbb{P}^2$ we may claim that it is the \textit{simplest situation} when the corresponding physical world is reach in properties. 

As far as systems with a phase space containing time derivatives of higher orders are concerned, they seem not to admit the introduction of the energy at all in a self-consistent way.  In order to explain this fact we reproduce the construction of the governing equations for such systems of particles using Lagrangian formulation of Newtonian mechanics based on the principle of least actions. It is worthy of noting that in some sense Lagrangian formulation of mechanics is more general than its formulation directly appealing to Newton's laws. Indeed in the latter case the existence of energy is an additional assumption imposing certain conditions on the forces with which physical particles interact with one another. In Lagrangian formulation the existence of some function, the Lagrangian $L$, reduced then to the system energy is the pivot point and the derived equation governing the system dynamics originally contain the forces meeting the required conditions.

\begin{figure}[t]
	\begin{minipage}[b]{0.49\columnwidth}
	\includegraphics[width=\columnwidth]{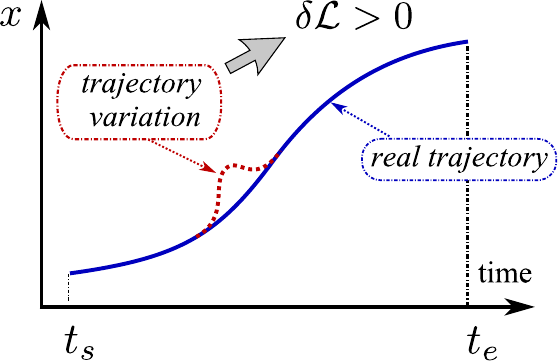}
	\end{minipage}\quad
	\begin{minipage}[b]{0.48\columnwidth}
	\caption{Illustration of the principle of least actions describing the minimality of the functional $\mathcal{L}\{x(t)\}$ \eqref{Red:10ad:1} taken at the real trajectory $\{x(t)\}$ with respect to its variations.}
	\label{Fig:Lagrang}
	\end{minipage}		
\end{figure}

As the general case, let us consider a system with the phase space
\begin{equation*}
\mathbb{P}^{m} = \{\theta_{m}\} =
 \left\{
x,\,\frac{dx}{dt},\,
\frac{d^2x}{dt^2},\,\ldots\,,\frac{d^{m-1}x}{dt^{m-1}}
\right\},
\end{equation*}
where $m$ is a certain number not necessary equal to 2. The pivot point of Lagrangian formalism is the introduction of a certain functional $\mathcal{L}\{x(t)\}$ determined for any arbitrary trajectory $\{x(t)\}_{t=t_s}^{t=t_e}$ starting and ending at some time moments $t=t_s$ and $t=t_e$, respectively. The notion of functional means that for any given trajectory $\{x(t)\}$ we can calculate a certain number  $\mathcal{L}\{x(t)\}$ which is treated as a measure of its ``quality'' in the realm of the Lagrangian mechanics. Since the systems at hand do not possess memory and cannot predict their future, all their significant characteristics including the ``quality'' of motion have to be determined by the local properties of the trajectory $\{x(t)\}$. In the given case, it is the collection $\theta_n$ of the phase variables $x,\, dx/dt,\,\ldots, d^{m-1}x/dt^{m-1}$. Therefore the functional $\mathcal{L}\{x(t)\}$ has to be of the integral form
\begin{equation}\label{Red:10ad:1}
\mathcal{L}\{x(t)\} = \int\limits_{t_s}^{t_e} L\left(
	x,\,\frac{dx}{dt},\,\frac{d^2x}{dt^2},\,\ldots\,,\frac{d^{m-1}x}{dt^{m-1}}
\right)dt\,,
\end{equation}
where $L(\ldots)$ is some function of these $n$ phase variables. The principle of least actions implies that ``the Nature chooses the best trajectories to implement'' the dynamics of mechanical systems. In other words, a real trajectory of system motion matches the minimum of functional~\eqref{Red:10ad:1} (or its maximum within the replacement $\mathcal{L}\to -\mathcal{L}$) with respect to all possible variations near this real trajectory (Fig.~\ref{Fig:Lagrang}). 

It is worthy of noting that in spite of its long-term history the fundamentality of the principle of least actions is up to now a challenging problem and there are a number of arguments for and against it from various points of view. Their brief review can be found, e.g., in \citet{stoltzner2006classical} as well as a detailed analysis of its ontological roots has be given by
\citet{stoltzner2003principle,stoltzner2009can,katzav2004dispositions,smart2015dispositions,terekhovich2015metaphisics}. Nevertheless its high efficient in many different branches of physics strongly argues for its real fundamentality. In Section~\ref{VFSSD} I have demonstrated that this principle can be derived based on the concepts of thick presentism for systems whose phase space contains hight order time derivatives as individual phase variables.

In the case of Newtonian mechanics, i.e., for systems with the phase space $\mathbb{P}^2$ the principle of least actions leads to the governing equations of type~\eqref{Red:5} admitting the introduction of the energy given by the expression
\begin{equation}\label{Red:10ad:2}
\mathcal{H}(x,v) = v \frac{\partial L(x,v)}{\partial v} - L(x,v)\,.
\end{equation}
In the general case (for $m>2$) it is also possible to construct some function having properties similar to that of energy,%
\footnote{An example of how to construct an ``energy,'' i.e., Hamiltonian for systems with the phase space $\mathbb{P}^3 = \{x,v,a\}$ has been demonstrated, e.g., by \citet{lubashevsky2003rational}.} 
however we meet the following challenging problem. 

Let $\{x(t)\}$ be a motion trajectory at which functional~\eqref{Red:10ad:1} attains its minimum (or maximum). Then for all small perturbations $x(t)+ \delta x(t)$ of this trajectory similar to one shown in Fig.~\ref{Fig:Lagrang} the variation $\delta\mathcal{L}=\mathcal{L}\{x(t)+\delta x(t)\}-\mathcal{L}\{x(t)\}$ of this functional has to be equal to zero in the linear approximation in $\delta x(t)$. It reads
\begin{multline}\label{m2prop:1}
\delta\mathcal{L} = \int\limits_{t_s}^{t_e}\left\{
     \frac{\partial L}{\partial x}\cdot \delta x(t) +
     \frac{\partial L}{\partial x^{(1)}}\cdot \delta\left[\frac{dx(t)}{dt}\right]  +
     \frac{\partial L}{\partial x^{(2)}}\cdot \delta\left[\frac{dx^2(t)}{dt^2}\right]  + \ldots\right.\\
     \left. {} +
     \frac{\partial L}{\partial x^{(m-1)}}\cdot \delta\left[\frac{dx^{m-1}(t)}{dt^{m-1}}\right] 
     \right\}dt = 0\,,
\end{multline}
where the symbol $x^{(p)}$ (with $p=1,2,\ldots m-1$) denotes the corresponding time derivative, $x^{(p)} = d^px(t)/dt^p$, treated as the argument of the function $L(x,x^{(1)},x^{(2)},\ldots, x^{(m-1)})$. Using the identities
\begin{gather*}
	 \delta\left[\frac{dx^p(t)}{dt^p}\right]= \frac{d^p [\delta x(t)]}{dt^p} = \frac{d}{dt} \left[\frac{d^{p-1}  [\delta x(t)]}{dt^{p-1}}\right] 
	 \quad(\text{in the latter case $p>1$}), \\
\intertext{the rule of integration by parts}
	  \int\limits_{t_s}^{t_e}  U(t) \frac{dV(t)}{dt}\, dt =  \Big[U(t)V(t)\Big]^{t=t_e}_{t=t_s} - \int\limits_{t_s}^{t_e}  V(t) \frac{dU(t)}{dt}\, dt \,,\\
\intertext{and choosing the trajectory perturbations $\delta x(t)$ such that (it is our right)}
      	  \delta x(t)|_{t=t_s, t_p} = 0\,, \qquad \left.\frac{d^p [\delta x(t)]}{dt^p}\right|_{t=t_s, t_p} = 0\quad(\text{for $p=1,2,\ldots,m-1$})
\end{gather*}
equality~\eqref{m2prop:1} is reduced to
\begin{multline}\label{m2prop:2}
\delta\mathcal{L} = \int\limits_{t_s}^{t_e}\left\{
\frac{\partial L}{\partial x} -
\frac{d}{dt}\,\frac{\partial L}{\partial x^{(1)}}  +
\left(\frac{d}{dt}\right)^2\frac{\partial L}{\partial x^{(2)}}  - \ldots\right.\\
\left. {} + (-1)^{m-1}
\left(\frac{d}{dt}\right)^{m-1}\frac{\partial L}{\partial x^{(m-1)}} 
\right\}\delta x(t)\,dt = 0\,.
\end{multline}
Because equality~\eqref{m2prop:2} must hold for any particular perturbation of the trajectory $\{x(t)\}$ this trajectory has to obey the equation
\begin{equation}\label{m2prop:3}
\frac{\partial L}{\partial x} -
\frac{d}{dt}\,\frac{\partial L}{\partial x^{(1)}}  +
\left(\frac{d}{dt}\right)^2\frac{\partial L}{\partial x^{(2)}}  - \ldots + (-1)^{m-1}
\left(\frac{d}{dt}\right)^{m-1}\frac{\partial L}{\partial x^{(m-1)}} 
=0\,.
\end{equation}
The time derivative of the highest order contained in equation~\eqref{m2prop:3} is 
$$d^{2(m-1)}x/dt^{2(m-1)}\,;$$ 
it enters this equation via the last term as the item
\begin{equation*}
(-1)^{m-1}\frac{\partial^2 L}{\partial [x^{(m-1)}]^2}\cdot \frac{d^{2(m-1)}x}{dt^{2(m-1)}}\,. 
\end{equation*}
When the derivative $\partial^2 L/\partial [x^{(m-1)}]^2$ is not equal to zero,\footnote{In the case of Newtonian mechanics with the phase space $\mathbb{P}^2$ the corresponding term  is just the mass $m$ of a given particle, $\partial^2 L/\partial [x^{(2)}]^2 = m$.} 
i.e., the Lagran\-gian $L(\ldots)$ is not a linear function with respect to its argument $d^{m-1}x/dt^{m-1}$, equation~\eqref{m2prop:3} can be directly resolved with respect to the derivative 
\[
\frac{d^{2(m-1)}x}{dt^{2(m-1)}}
\]
and rewritten as
\begin{equation}\label{Red:10ad:3}
      \frac{d^{2(m-1)}x}{dt^{2(m-1)}} = \Phi^*\left(
      x,\,\frac{dx}{dt},\,\frac{d^2x}{dt^2},\,\ldots\,,\frac{d^{2m-3}x}{dt^{2m-3}}
      \right),
\end{equation}
where $\Phi^*(\ldots)$ is a certain function. Equation~\eqref{Red:10ad:3} governs the dynamics of the system in issue and can be regarded as its basic law written in the form of differential equation.\footnote{Lagrangian mechanics with higher order time derivatives is well known and was developed during the middle of the XIX century by \cite{ostrogradski1850}. So here I have presented the results in a rather symbolic form emphasizing the features essential for our consideration. Mechanics dealing with equation~\eqref{Red:10ad:3} with $m>2$ as one describing some initial value problem faces up to the Ostrogradsky instability \citep[see, e.g.,][]{woodard2007instability,stephen2008ostrogradski,smilga2009comments}, which can be used for explaining why no differential equations of higher order than two appear to describe physical phenomena \citep{motohashi2015third}. There are also arguments for the latter conclusion appealing to metaphysical aspects of time changes in physical quantities \citep{easwaran2014physics}.}

When the number of the phase variables forming the phase space $\mathbb{P}^m$ is larger than two, $m>2$, the obtained governing equation~\eqref{Red:10ad:3} comes in conflict with the initial assumption about the properties of the given physical system. The matter is that the number of the arguments of the function $\Phi^*(\ldots)$ \textit{exceeds} the dimension of the phase space $\mathbb{P}^m$ because $2(m-1) > m$ for $m > 2$. Thereby we cannot treat equation~\eqref{Red:10ad:3} as a law governing the deterministic motion of a certain  dynamical system in the phase space  $\mathbb{P}^m$. Indeed, to do this we need that the point $\theta_m = \{x,\, dx/dt,\,\ldots, d^{m-1}x/dt^{m-1}\}$ of the phase space $\mathbb{P}^m$ determine completely the rate of the system motion in it, in other words, the corresponding governing equation should be of form $d\theta_m/dt = \Phi(\theta_m)$ (see equation~\eqref{Red:5}). However, the obtained equation~\eqref{Red:10ad:3} stemming from the principle of least actions contradicts this requirement because its right hand side contains the time derivatives higher than $d^{(m-1)}x/dt^{(m-1)}$ for $m>2$. Therefore in order to construct a solution of equation~\eqref{Red:10ad:3} dealing with only the phase space $\mathbb{P}^m$ we need some additional information about, for example, the terminal point of the analyzed trajectory. The latter feature, however, contradicts the principle of microlevel reducibility because according this principle the current state of such a system should determine its further motion completely.

Summarizing this discussion we see that only in the case of $m=2$ equation~\eqref{Red:10ad:3} following from the principle of least actions for trajectories in the phase space $\mathbb{P}^m$ admits the interpretation in terms of a certain dynamical system whose motion is \textit{completely} specified within this phase space. If $m=1$ the minimality of functional~\eqref{Red:10ad:1} does not describe any dynamics. Therefore, it is likely that the notion of the energy $\mathcal{H}(x,v)$ can be introduced in a self-consistent way only for systems with the phase space $\mathbb{P}^2=\{x,v\}$ where the governing laws can be written as differential equations of the second order.

\section{Conclusion}\label{conclusion:physsyst}

In the given paper I have presented arguments for the relationship between the notions and formalism used in the basic laws of classical physics and the existence of the microlevel of description of the corresponding physical systems which obeys the principle of microlevel reducibility.

According to this principle, first, any system belonging to the realm of classical physics admits the representation as an ensemble of structureless particles with certain properties ascribed to them individually. Second, the interaction between these particles is supposed to be determined completely by their individual properties and to meet the superposition principle. In the given case the superposition principle is reduced either to (\textit{i}) the model of long-distant pair-wise interaction between particles or (\textit{ii})  the model of local interaction between particles and some fields with linear properties. Within the former model  the dynamics of any system is determined completely by the current values taken by the individual properties of its particles. Within the latter model this statement also holds provided the particle properties instantiated withing some time interval are taken into account. 

\vspace*{0.25\baselineskip}

Whence I have drawn the following conclusions, where, for short, the particle long-distant interaction is implied to be the case if a specific model of particle interaction is not noted explicitly.   
\begin{itemize}
	\item Laws governing the dynamics of such systems can be written within the formalism of ordinary differential equations dealing with time derivatives of the particle's individual properties. It has been justified (\textit{i}) appealing to the concepts of thick presentism regarding the flow of time as a sequence of bold instants and (\textit{ii}) introducing two dimensional time---the physical time and the meta-time.    
	\vspace*{0.25\baselineskip}
	
	\item There is a limit in the two-dimensional time dynamics called the limit of steady-state laws that admits the introduction of governing laws describing the system motion in the form of differential equations relating the time derivatives of order less than a certain integer $m$ to the time derivative of order $m$. Moreover in this case it is possible to introduce the notion of Lagrangian---a function depending on the time derivatives of order less than $m$. Its integral over a trial trajectory takes the minimum at the real trajectory, which justifies the principle of least actions well known in physics.     
	\vspace*{0.25\baselineskip}

	\item Dynamics of such systems can be described as their motion in the corresponding phase space $\mathbb{P}^m$. A point of this phase space admits interpretation as a collection of all time derivatives of the particles' individual properties whose order is less than a certain integer $m$; naturally, this collection includes these properties too.  The position of a given system in its phase space $\mathbb{P}^m$ determines the rate of system motion in this phase space, which enables us to introduce the notion of initial conditions and the concept of determinism of physical systems. 
	\vspace*{0.25\baselineskip}
	
	\item The energy conservation should be a consequence of some general laws governing various systems of a given type rather than particular circumstances. Within this requirement, for a system to admit the introduction of energy, its phase space $\mathbb{P}^2$ must comprise only the individual properties of the constituent particles and the corresponding time derivatives of the fist order. The dynamics of such systems is described by differential equations of the second order with respect to time derivatives, which is exactly the case of Newtonian mechanics. In these sense the systems belonging to the realm of classical physics take the unique position among the other plausible models. 
	\vspace*{0.25\baselineskip}
	
\end{itemize}

\end{document}